\documentclass{article}

\usepackage{microtype}
\usepackage{graphicx}
\usepackage{subfigure}
\usepackage{booktabs} 

\usepackage{hyperref}

\usepackage[accepted]{icml2024}

\usepackage{xcolor}

\usepackage{amsmath}
\usepackage{amssymb}
\usepackage{mathtools}
\usepackage{amsthm}

\usepackage[capitalize,noabbrev]{cleveref}

\theoremstyle{plain}

\theoremstyle{definition}

\theoremstyle{remark}

\usepackage[textsize=tiny]{todonotes}

\DeclareMathOperator*{\argmax}{arg\,max}

\icmltitlerunning{Explaining Agent Behavior to a Human Terminator}

\begin{document}

\twocolumn[
\icmltitle{``Trust me on this'' \\  Explaining Agent Behavior to a Human Terminator}



\icmlsetsymbol{equal}{*}

\begin{icmlauthorlist}
\icmlauthor{Uri Menkes}{technion}
\icmlauthor{Assaf Hallak}{nvidia}
\icmlauthor{Ofra Amir}{technion}

\end{icmlauthorlist}

\icmlaffiliation{technion}{Faculty of Data and Decision Sciences, Technion, Israel Institute of Technology}
\icmlaffiliation{nvidia}{NVIDIA Research, Israel}

\icmlcorrespondingauthor{Uri Menkes}{urimenkes@campus.technion.ac.il}

\icmlkeywords{Machine Learning, ICML}

\vskip 0.3in
]



\printAffiliationsAndNotice{}  

\begin{abstract}
Consider a setting where a pre-trained agent is operating in an environment and a human operator can decide to temporarily terminate its operation and take-over for some duration of time. These kind of scenarios are common in human-machine interactions, for example in autonomous driving, factory automation and healthcare. In these settings, we typically observe a trade-off between two extreme cases -- if no take-overs are allowed, then the agent might employ a sub-optimal, possibly dangerous policy. Alternatively, if there are too many take-overs, then the human has no confidence in the agent, greatly limiting its usefulness. In this paper, we formalize this setup and propose an explainability scheme to help optimize the number of human interventions. 
\end{abstract}

\section{Introduction}



With the growing deployment of AI, it is becoming increasingly common for humans and AI to collaborate on tasks. In this context, the agent and human work together as a team, combining their respective skills and knowledge to achieve a shared goal. Here, we focus on a human-AI collaboration setting where an AI agent operates autonomously, and a human operator can intervene when needed. Specifically, we study sequential decision-making settings where the agent's policy is typically learned using reinforcement learning (RL). For instance, a human driver may take over an autonomous vehicle, while a factory operator may intervene if a robot gets stuck. This setting is similar to the ``termination'' framework  \citep{tennenholtz2022reinforcement} in which at every step, a terminator may terminate the interaction as a function of a mistrust function, adjusted according to the observed trajectory. 

A key challenge in human-AI teamwork is achieving complementary team performance, where the team outperforms the human or the AI operating alone~\citep{bansal2021does}. In the context of the termination setting, the human will need to have a good understanding of the agent's policy to intervene optimally. If the human cannot correctly predict the agent's actions, she may intervene too little (resulting in lower team utility) or intervene unnecessarily (resulting in a higher cost).  

Our work focuses on developing a methodology for explaining AI to human users in a Human-AI team to enhance the team's performance. We hypothesize that providing users with explanations and descriptions of the agent's policy can increase users' appropriate trust and thus improve their ability to correctly identify scenarios in which they should intervene in the agent's decision-making. However, understanding the decision-making of RL agents is challenging, as they operate in large state spaces and their decision-making is affected by long-term outcomes.

Prior work on explainable RL proposed policy summarization as a method to convey the global policy of an agent to a user. In this approach, the user is presented with a set of trajectories demonstrating the behavior of an agent in selected world-states~\citep{amir2018agent}. We aim to create a policy summary method tailored to the termination setting, which can help users determine in which scenarios the agent does not act optimally and intervene accordingly. To do so, we formalize the proposed setup and explain the required modules to solve this task. We provide initial solutions to some of the challenges and propose an evaluation procedure for human-subject experiments. 

\begin{figure*}[ht!]
    \centering    
    \includegraphics[width=0.9\textwidth]{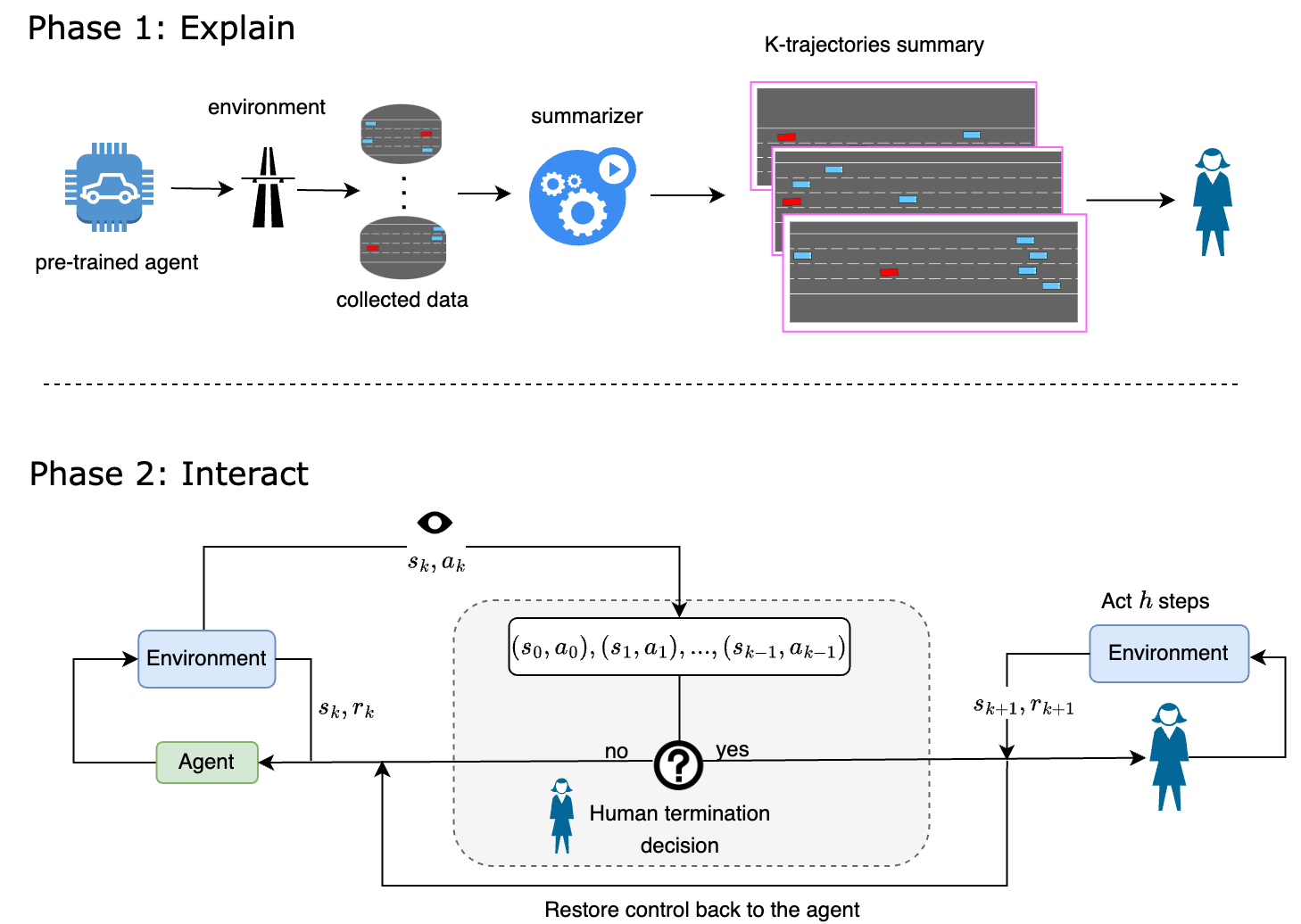}
    \caption{\textbf{Top. }To explain the agent's behavior to the human, we use a summarizer module to select a subset of trajectories from a large set of simulated data. The selected subset is rendered and shown to the human to improve her understanding of the agent. \textbf{Bottom. }During online interaction, the human can decide at any point to terminate the agent and take-over the policy for $h$ steps.}
    \vspace{-0.3cm}
    
    \label{fig:setup}
\end{figure*}

\section{Related Work}

\textbf{Human-AI Teamwork.} Prior work on human-AI teamwork has shown that achieving complementary team performance is challenging, and that often using either the AI or the human alone results in better performance. While some experiments showed that if a user develops an accurate mental model of the AI's errors, better complementarity can be achieved~\citep{bansal2019beyond}, other experiments showed that adding explanations for the AI recommendations did not always help users better utilize the AI~\citep{bansal2021does}. In this work, we study the effect of adding policy explanations in a human-AI teamwork setting where an RL agent can act autonomously but the human operator can intervene and terminate it if she expects the agent's behavior to be sub-optimal. 

\textbf{Explainable reinforcement learning.} Various approaches to explainable RL (XRL) have been proposed in the literature~\citep{milani2023explainable}. \emph{Local} explanations provide information regarding the agent's decision-making in a particular world-state, e.g., by showing a saliency map of the agent's attention~\citep{greydanus2018visualizing}. However, in the termination setting it may not be feasible to explain each state in real-time and expect the user to decide immediately whether to intervene. Therefore, we focus on \emph{global} explanations that aim to describe the overall policy of the agent~\citep{amitaisurvey}. Global explanation approaches include interpretable representations of the policy (e.g., learning a decision-tree mimicking the policy), and policy summarization methods~\citep{amir2018agent} which aim to describe the policy of the agent by demonstrating its behavior in selected world-states.
Prior approaches to policy summarization have constructed summaries based on criteria such as state importance, diversity, and state visitation frequency~\citep{amir2018highlights,huang2019enabling,sequeira2020interestingness}. In this work, we develop new criteria for policy summaries tailored to the termination settings and aim to evaluate the usefulness of such explanations in these settings.

\section{Background}\label{sec:background}

The RL framework \citep{sutton2018reinforcement} is commonly described formally using an MDP which is a tuple $\left( \mathcal{S},\mathcal{A}, R, P, \rho, \gamma \right)$, such that: $\mathcal{S}$ is the set of states; $\mathcal{A}$ is the set of actions; $R$ is a reward function $R:\mathcal{S} \times \mathcal{A} \rightarrow \mathbb{R}$ assigning the immediate reward $R(s,a)$ for taking action $a$ in state $s$; $P$ is a transition probability function $P(s|a,s') \rightarrow [0,1]$ s.t  $s, s' \in S, a \in A$;  $\rho$ is the distribution over initial states, and  $\gamma$ is the discount factor.




In the beginning of an episode, the environment starts from $s_0 \sim \rho$, then the agent selects an action $a_0\in\mathcal{A}$. The environment transitions to the next state $s_1 \sim P(s'|s_0, a_0)$, returns a reward $R(s_0, a_0)=R_1$ and continues to the next time step. At time step $t$ the environment returns a state $s_t$ and a reward $R_t$ and the agent takes action $a_t$ resulting in $R(s_t, a_t)=R_{t+1}$. 

The goal of the agent is to maximize its expected cumulative discounted reward over time by employing a policy $\pi.$ Given $\pi$, a \emph{value function} assigns a numerical value to a state:
$V^\pi (s) = \mathbb{E}^\pi[\sum_{k=0}^{\infty}\gamma^{k}R_{t+k+1} | s_t=s].$ Similarly, an \emph{action-value function} (or \emph{Q-function}), assigns a value to any state-action pair: $Q^{\pi}(s,a)=\mathbb{E}^{\pi}[\sum_{k=0}^{\infty}\gamma^{k}R_{t+k+1}| s_t=s,a_t=a].$ Solving an RL problem defined as above means finding an optimal policy in terms of the $Q$ (or $V$) function, i.e. $
    \pi^* = \underset{\pi}{\argmax}Q^\pi(s,a) \ \forall \quad s \in \mathcal{S},a \in \mathcal{A}.$

\section{Setup}
In a recent paper, \citet{tennenholtz2022reinforcement} presented an extension to the traditional MDP framework, in which an external non-Markovian observer oversees the agent in action and may terminate its interaction with the environment. We propose a similar setup that mimics more closely the real world where the human is not part of the training loop, but instead needs to operate a given fixed policy agent. Instead of modeling directly the human's trust in the agent, we aim to improve the human understanding of the agent performance in different areas in the state space, so she can make an informed decision on when to terminate and take-over.

We define the Termination Markov Decision Process (TerMDP) by the tuple $\mathcal{M}_T=\left(\mathcal{S},\mathcal{A},R,P,\rho, \gamma,c, h, \pi^\text{human}, \pi^{\text{agent}}\right)$, where $\mathcal{S},\mathcal{A},R,P, \rho, \gamma$ are as defined in Section \ref{sec:background}, with the addition of the termination cost function $c \in \mathcal{R},$ the take-over horizon $h\in\mathbb{N},$ and the human and agent policies $\pi^\text{human}, \pi^{\text{agent}}$, correspondingly. In each time step, after observing the current state, the human can decide to terminate the agent and apply $\pi^{\text{human}}$ for $h$ steps. In this case, we assume a cost of $c$ is incurred. A diagram of the proposed setup is given in Figure~\ref{fig:setup} (bottom). 

\textbf{Human and AI policies. }In the proposed setup we assume both policies $\pi^\text{human}, \pi^\text{agent}$ are fixed. In some real-world cases, both the agent and the human employ sub-optimal policies: the agent due its model capacity and limited data (for example in a sim2real scenario; \cite{hofer2021sim2real}), and the human due to problem complexity and partial observability. In other cases, both policies are optimal, but reflect different reward functions stemming from the difficulty of reward engineering to match human preferences. 

\textbf{Termination policy. }Since the problem is well defined, given all parameters it is possible to find an optimal termination policy which optimizes the overall cumulative reward given the termination cost and horizon. To solve it, we can consider the take-over to be an option \citep{precup1998theoretical} and employ existing solutions. Since the agent's policy is usually unknown to the human, such solutions are non-trivial in practice. We leave further insights on this direction to future research.

\subsection{Problem Definition}
To improve the team's performance, i.e., to maximize the expected utility, the human should understand the agent's decision process, anticipate its actions and, more importantly for our goal, know where it would need their help (by taking over). This is done by showing the human an explanation about the agent's policy. 

To this end, we decompose the problem into two phases. In phase 1, we explain the agent to the human. To do so, we propose to collect many trajectories and summarize them into a subset of trajectories that exhibit the agent's behavior in different situations. The human then sees the explanation, and the next phase can start. In phase 2, we follow the TerMDP setup described above, where now the human can make an informed decision about the environment states that require termination. Both phases are illustrated in Figure~\ref{fig:setup}. Subsequently, we reduce the human-AI termination problem to that of summarization, i.e., which subset of trajectories should we exhibit to the human, such that her future interaction with the agent will be composed of beneficial terminations. 

\begin{figure*}[ht!]
    \centering    
    \includegraphics[width=\textwidth]{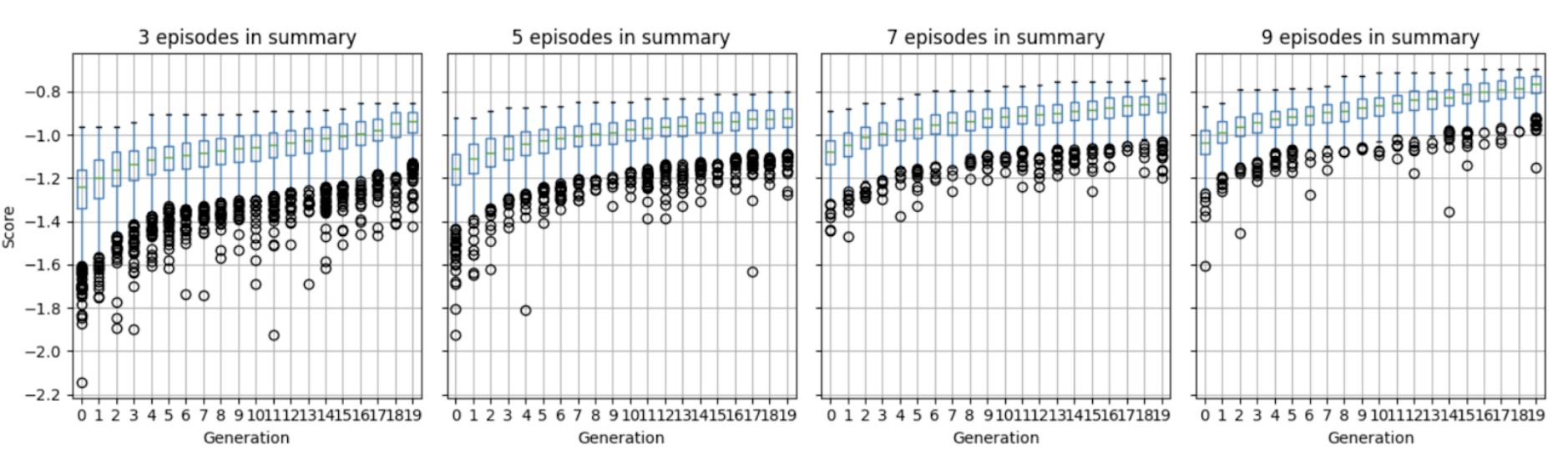}
    \caption{The genetic algorithm optimizes for score \ref{s3} and is tested across different summary lengths. The score demonstrates improvement as the generation size increases. Similar convergence trend results were observed for scores \ref{s1}, \ref{s2}}.
        \vspace{-0.2cm}
    \label{fig:genetic}
\end{figure*}

\subsection{Algorithmic Approach}
Our approach involves extracting the summary from data collected through the agent's past performance. Specifically, we run a trained agent in its environment following its policy, $\pi^{\text{agent}}$, collecting data on trajectories (i.e., a series of transitions) of length $N$, denoted by $D$ (i.e., $\tau \in D, \tau=\{(s^\tau_0,a^\tau_0)...(s^\tau_n,a^\tau_n)\})$. We then evaluate candidate summaries denoted $\{S_j \subset D \}$, each consisting of a subset of $K$ trajectories. We extract the optimal summary $S^{*}$ from this data by optimizing a score which aims to characterize a good summary, discussed next. 

\textbf{What is a good summary for termination? } Several optimization criteria for policy summaries have been proposed in the literature. One approach by \cite{lage2019exploring} suggests that a good summary should minimize the reconstruction error of the agent policy over the entire state-space (weighted by the induced distribution over trajectories). However, in the termination setup, that might be sub-optimal. For instance, consider an AV driving on an open road with no other vehicles in its vicinity. Even if the AV acts sub-optimally by sporadically changing lanes, this has little effect on the driver as it does not affect the riding time or safety of the travel. Thus, even if an open road is a fairly common scenario in the data, the information on the policy there is inconsequential for termination. Alternatively, we might prefer to show summaries that focus on situations characterized by high uncertainty in the outcome of different actions. 

We propose several scoring schemes to evaluate a candidate summary $S_j$. Some of the proposed scores are aimed at reconstruction while others are aimed at reducing uncertainty.  
For the first two scoring schemes, we employ ensemble training to estimate the Q-function of the agent. Initially, multiple networks undergo training using a deep Q-learning algorithm (DQN) on the entirety of the collected set of episodes $D$. This ensemble approximates the agent's learned Q-values (which we do not assume access to), denoted by $Q^{\text{agent}}$. For each candidate summary $S_j$, we train another ensemble, $\hat{Q}_j$, that uses only the trajectories in $S_j$ to approximate the agent's Q-values. To estimate the scores, we use a held-out set of trajectories $\Tilde{D}$ to maintain statistical independency.


\paragraph{Score 1: reducing uncertainty.} Our first score is:
\begin{equation}\label{s1}
     \text{Score($S_j$)}=\frac{1}{|\Tilde{D}|}\sum_{i \in \Tilde{D}}\frac{ Var (\hat{Q}_j(s_0^i, a_0^i) ) }{ Var(Q^{\text{agent}}(s_0^i, a_0^i) ) },
\end{equation}
where the variance is taken with respect to the ensemble. We hypothesize that complex scenarios lead to high variance in the denominator. Therefore, an effective summary should encompass such scenarios, reducing the variability of the reconstructed Q-function (numerator) and resulting in a lower score.

\paragraph{Score 2: minimizing Q-value reconstruction error.} This score was devised to reflect the true agent's policy, aiming to avoid edge cases that could lead to significant differences in the averages of the ensembles. It is computed as follows:
 
\begin{equation}\label{s2}
    \text{Score($S_j$)}=\frac{1}{|\Tilde{D}|}\sum_{i \in \Tilde{D}} \left(\mathbb{E}[Q^{\text{agent}}(s_0^i, a_0^i)] - \mathbb{E}[\hat{Q}(s_0^i, a_0^i)]\right)^2,
\end{equation}
where the expectation is taken w.r.t the ensemble. 

\paragraph{Score 3: maximizing policy likelihood.}
For a third score, we adopt a different approach, employing a classification model that directly imitates the policy instead of an ensemble estimating the Q values. We hypothesize this approach may better align with how a human user would interpret the summary, akin to imitation learning \citep{hussein2017imitation}. A high likelihood over an unseen set of states can be interpreted as a solid understanding of the agent's policy.

For each candidate summary $S_j$, we train a classification model (XGBoost), denoted as $f_j$, on the data extracted from the summary: the states serve as input features, while the actions are the output classes. Subsequently, we determine the score as the average-log-likelihood of the model across the evaluation set. Formally:
\begin{equation}\label{s3}
    \text{Score($S_j$)}=\frac{1}{|\Tilde{D}|} \sum_{(s,a) \in \Tilde{D}} \log (P_{f_j}(a|s))
\end{equation}

\subsection{Optimizing the score}
Summary optimization requires going over all $K$-subsets of a given large set. This optimization is naturally combinatorical making exhaustive search infeasible. To tackle this problem we propose to 
employ a genetic algorithm that receives as input the dataset $D$, the desired summary size $K$, and the scoring scheme to optimize for. Starting from a randomized population of subsets, in each optimization generation we generate several entirely new samples, and also generate new subsets by randomly taking  trajectories from two subsets with good scores. We also leave in the population some of the subsets with good scores for future sample generation.

\begin{figure}[ht]
\vskip 0.2in
\begin{center}
\centerline{\includegraphics[width=0.6\columnwidth]{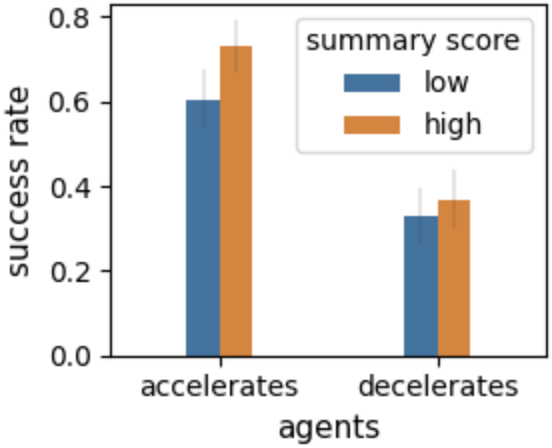}}
\caption{Results of the termination task study. 
Participants who viewed the higher score summaries outperformed participants who viewed the lower score summaries ($n=160$, $p$-value = 0.02). }
\label{fig:results}
\end{center}
\vskip -0.2in
\end{figure}

\subsection{Experimental Methodology}
Our approach involves training agents and evaluating our explanation methods in simulated environments, with the ability to fine-tune as needed. The environment we simulate is \textit{Highway-Env} \citep{highway-env}, where the agent operates as a car on a highway. We utilize the DQN algorithm to train an agent, gather data (a set of episodes) from it, and extract summaries optimized for each of the scores mentioned earlier. The convergence of the genetic algorithm for various summary sizes is depicted in Figure~\ref{fig:genetic}.

We conducted an initial pilot proxy user study to validate our explanation methods. Participants were presented with summaries of an agent at different score levels (maximum, median, or low score). They were then asked to identify the summarized agent among others with notably different policies. Our preliminary findings indicate that humans exhibit a bias towards collisions. Specifically, a summary containing collisions was more likely to lead participants to associate it with scenarios involving collisions, irrespective of speed profiles and lane changes. This suggests that people's reasoning is more nuanced and scores may need to be integrated with additional heuristics.


Next, we conducted a user study to test our explanation methods and assess the effectiveness of various scoring schemes for the termination task. The experiment focused on presenting users with agent summaries based on Score \ref{s3}, we leave experiments on the other scores to future work. To create a scenario where an agent is sub-optimal in some region of the state-space, we implemented an agent that shifts between two policies: one extensively trained, and the other simplified, focusing on a single action (either acceleration or deceleration). During execution, the primary ``high-quality'' policy is implemented across three out of four lanes, while the ``degenerate'' policy is applied in the leftmost lane. Human intervention is needed to improve the flawed agent's performance in the leftmost lane. Participants were presented with a summary of an agent with either a high or low score, consisting of $K=5$ episodes. This length was chosen based on feedback from the proxy study, where participants indicated that a 7-episode summary felt too long, and considering the convergence results (see Figure \ref{fig:genetic}). After reviewing the summary, participants were presented with a series of 10 questions. Each question included a pair of scenarios showing the agent acting in the environment, which is paused after a certain time duration. Participants had to decide in which of the two scenarios they would choose to terminate the agent's operation and assume manual control of the car. Five of the questions had a ground truth, pausing the scenario where the agent was about to act sub-optimally. The other five questions were control questions with no clear correct answer, and their results were not considered in the analysis.
As expected, users who were shown an optimized summary performed better than those who viewed a lower score summary and were more likely to identify left-lane behavior as requiring intervention (Figure \ref{fig:results}).

\section{Discussion and Future Work}
In this paper, we present the problem of supporting a human operator in deciding whether and when to terminate an agent's operation and assume control. We propose to achieve this by presenting users with a summary of the agent's policy, optimized for this specific human-AI teamwork scenario. To this end, we propose three alternative scoring schemes optimized for different aspects of policy reconstruction. Initial user study results suggest that our proposed methods may be effective for this task. Future work will evaluate the impact of our approach on users' ability to collaborate effectively with agents in more realistic experiments. While promising results were presented in the termination study, some findings from the proxy study indicated significant differences in how users mentally model the agent's policy and their understanding of an \textit{optimal} human policy within the driving domain. To address this, we plan to expand the study to explore other domains as well.

\paragraph{Acknowledgments.}
The research was funded by the European Research Council (grant \#101078158 - CONVEY).


\begin{thebibliography}{16}
\providecommand{\natexlab}[1]{#1}
\providecommand{\url}[1]{\texttt{#1}}
\expandafter\ifx\csname urlstyle\endcsname\relax
  \providecommand{\doi}[1]{doi: #1}\else
  \providecommand{\doi}{doi: \begingroup \urlstyle{rm}\Url}\fi

\bibitem[Amir \& Amir(2018)Amir and Amir]{amir2018highlights}
Amir, D. and Amir, O.
\newblock Highlights: Summarizing agent behavior to people.
\newblock In \emph{Proceedings of the 17th International Conference on Autonomous Agents and MultiAgent Systems}, pp.\  1168--1176, 2018.

\bibitem[Amir et~al.(2018)Amir, Doshi-Velez, and Sarne]{amir2018agent}
Amir, O., Doshi-Velez, F., and Sarne, D.
\newblock Agent strategy summarization.
\newblock In \emph{Proceedings of the 17th International Conference on Autonomous Agents and MultiAgent Systems}, pp.\  1203--1207, 2018.

\bibitem[Amitai \& Amir(2023)Amitai and Amir]{amitaisurvey}
Amitai, Y. and Amir, O.
\newblock A survey of global explanations in reinforcement learning.
\newblock In \emph{Explainable Agency in Artificial Intelligence}, pp.\  21--42. CRC Press, 2023.

\bibitem[Bansal et~al.(2019)Bansal, Nushi, Kamar, Lasecki, Weld, and Horvitz]{bansal2019beyond}
Bansal, G., Nushi, B., Kamar, E., Lasecki, W.~S., Weld, D.~S., and Horvitz, E.
\newblock Beyond accuracy: The role of mental models in human-ai team performance.
\newblock In \emph{Proceedings of the AAAI conference on human computation and crowdsourcing}, volume~7, pp.\  2--11, 2019.

\bibitem[Bansal et~al.(2021)Bansal, Wu, Zhou, Fok, Nushi, Kamar, Ribeiro, and Weld]{bansal2021does}
Bansal, G., Wu, T., Zhou, J., Fok, R., Nushi, B., Kamar, E., Ribeiro, M.~T., and Weld, D.
\newblock Does the whole exceed its parts? the effect of ai explanations on complementary team performance.
\newblock In \emph{Proceedings of the 2021 CHI Conference on Human Factors in Computing Systems}, pp.\  1--16, 2021.

\bibitem[Greydanus et~al.(2018)Greydanus, Koul, Dodge, and Fern]{greydanus2018visualizing}
Greydanus, S., Koul, A., Dodge, J., and Fern, A.
\newblock Visualizing and understanding atari agents.
\newblock In \emph{International conference on machine learning}, pp.\  1792--1801. PMLR, 2018.

\bibitem[H{\"o}fer et~al.(2021)H{\"o}fer, Bekris, Handa, Gamboa, Mozifian, Golemo, Atkeson, Fox, Goldberg, Leonard, et~al.]{hofer2021sim2real}
H{\"o}fer, S., Bekris, K., Handa, A., Gamboa, J.~C., Mozifian, M., Golemo, F., Atkeson, C., Fox, D., Goldberg, K., Leonard, J., et~al.
\newblock Sim2real in robotics and automation: Applications and challenges.
\newblock \emph{IEEE transactions on automation science and engineering}, 18\penalty0 (2):\penalty0 398--400, 2021.

\bibitem[Huang et~al.(2019)Huang, Held, Abbeel, and Dragan]{huang2019enabling}
Huang, S.~H., Held, D., Abbeel, P., and Dragan, A.~D.
\newblock Enabling robots to communicate their objectives.
\newblock \emph{Autonomous Robots}, 43:\penalty0 309--326, 2019.

\bibitem[Hussein et~al.(2017)Hussein, Gaber, Elyan, and Jayne]{hussein2017imitation}
Hussein, A., Gaber, M.~M., Elyan, E., and Jayne, C.
\newblock Imitation learning: A survey of learning methods.
\newblock \emph{ACM Computing Surveys (CSUR)}, 50\penalty0 (2):\penalty0 1--35, 2017.

\bibitem[Lage et~al.(2019)Lage, Lifschitz, Doshi-Velez, and Amir]{lage2019exploring}
Lage, I., Lifschitz, D., Doshi-Velez, F., and Amir, O.
\newblock Exploring computational user models for agent policy summarization.
\newblock In \emph{IJCAI: proceedings of the conference}, volume~28, pp.\  1401. NIH Public Access, 2019.

\bibitem[Leurent(2018)]{highway-env}
Leurent, E.
\newblock An environment for autonomous driving decision-making.
\newblock \url{https://github.com/eleurent/highway-env}, 2018.

\bibitem[Milani et~al.(2023)Milani, Topin, Veloso, and Fang]{milani2023explainable}
Milani, S., Topin, N., Veloso, M., and Fang, F.
\newblock Explainable reinforcement learning: A survey and comparative review.
\newblock \emph{ACM Computing Surveys}, 2023.

\bibitem[Precup et~al.(1998)Precup, Sutton, and Singh]{precup1998theoretical}
Precup, D., Sutton, R.~S., and Singh, S.
\newblock Theoretical results on reinforcement learning with temporally abstract options.
\newblock In \emph{Machine Learning: ECML-98: 10th European Conference on Machine Learning Chemnitz, Germany, April 21--23, 1998 Proceedings 10}, pp.\  382--393. Springer, 1998.

\bibitem[Sequeira \& Gervasio(2020)Sequeira and Gervasio]{sequeira2020interestingness}
Sequeira, P. and Gervasio, M.
\newblock Interestingness elements for explainable reinforcement learning: Understanding agents' capabilities and limitations.
\newblock \emph{Artificial Intelligence}, 288:\penalty0 103367, 2020.

\bibitem[Sutton \& Barto(2018)Sutton and Barto]{sutton2018reinforcement}
Sutton, R.~S. and Barto, A.~G.
\newblock \emph{Reinforcement learning: An introduction}.
\newblock MIT press, 2018.

\bibitem[Tennenholtz et~al.(2022)Tennenholtz, Merlis, Shani, Mannor, Shalit, Chechik, Hallak, and Dalal]{tennenholtz2022reinforcement}
Tennenholtz, G., Merlis, N., Shani, L., Mannor, S., Shalit, U., Chechik, G., Hallak, A., and Dalal, G.
\newblock Reinforcement learning with a terminator.
\newblock \emph{Advances in Neural Information Processing Systems}, 35:\penalty0 35696--35709, 2022.

\end{thebibliography}
\bibliographystyle{icml2024}




\end{document}